\documentclass[10pt]{article}

\usepackage{doublespace}
\usepackage{note}
\usepackage{epsfig}
\usepackage{latexsym}

\begin{document}
\title{Choi's Proof and Quantum Process Tomography}
\author{Debbie W. Leung} 
\address{IBM T.J.~Watson Research Center, P.O.~Box 218, 
         Yorktown Heights, NY 10598, USA \\[1.5ex]}
\maketitle
\centerline{\today}

\vspace*{-4ex} 

\begin{abstract}

Quantum process tomography is a procedure by which an unknown quantum
operation can be fully experimentally characterized.
We reinterpret Choi's proof of the fact that any completely positive
linear map has a Kraus representation [Lin. Alg. and App., 10, 1975] 
as a method for quantum process tomography.
Furthermore, the analysis for obtaining the Kraus operators are
particularly simple in this method.

\end{abstract}

\section{Introduction}

The formalism of quantum operation can be used to describe a very
large class of dynamical evolution of quantum systems, including 
quantum algorithms, quantum channels, noise processes, and measurements.
The task to fully characterize an unknown quantum operation ${\cal E}$
by applying it to carefully chosen input state(s) and analyzing the
output is called quantum process tomography.
The parameters characterizing the quantum operation are contained in 
the density matrices of the output states, which can be measured using 
quantum state tomography~\cite{ADD}.
Recipes for quantum process tomography have been proposed
\cite{Poyatos97a,Chuang96d,DAriano98,DAriano00,Leung00t}. 
In earlier methods \cite{Poyatos97a,Chuang96d,DAriano98}, ${\cal
E}$ is applied to different input states each of exactly the input
dimension of $\cal E$.
In \cite{DAriano00,Leung00t}, ${\cal E}$ is applied to part of a fixed
bipartite entangled state.  In other words, the input to $\cal E$ is
entangled with a reference system, and the joint output state is
analyzed.

Quantum processing tomography is an essential tool in reliable quantum
information processing, allowing error processes and possibly
imperfect quantum devices such as gates and channels to be
characterized.
The method in~\cite{Chuang96d} has been experimentally demonstrated
and used to benchmark the fidelities of teleportation \cite{Nielsen99}
and the gate {\sc cnot} \cite{Childs01qpt}, and to investigate the
validity of a core assumption in fault tolerant quantum computation
\cite{Childs01qpt}.

The number of parameters characterizing a quantum operation $\cal E$,
and therefore the experimental resources for any method of quantum
process tomography, are determined by the input and output dimensions
of $\cal E$.  
However, different methods can be more suitable for different physical
systems.  
Furthermore, each method defines a procedure to convert the measured
output density matrices to a desired representation of $\cal E$, and
a simpler procedure will enhance the necessary error analysis.   

In this paper, we describe in detail the method initially reported in
\cite{Leung00t}, which is derived as a simple corollary of a
mathematical proof reported in \cite{Choi75a}.
Our goal is two-fold.  We hope to make this interesting proof more
accessible to the quantum information community, as well as to provide
a simple recipe for obtaining the Kraus operators of the unknown
quantum operation.
In the rest of the paper, we review the different approaches of
quantum operations, describe Choi's proof and the recipe for quantum
process tomography in Sections \ref{sec:3approachs}, \ref{sec:proof}, and
\ref{sec:qpt}.  We conclude with some discussion in
Section \ref{sec:conclude}.

\section{Equivalent approaches for quantum operations}
\label{sec:3approachs}

A quantum state is usually described by a density matrix $\rho$ that
is positive semidefinite ($\rho \geq 0$, i.e., all eigenvalues are
nonnegative) with $\tr(\rho) = 1$.  
A quantum operation $\cal E$ describes the evolution of one state
$\rho$ to another $\rho' = {\cal E}(\rho)$.  

More generally, let ${\cal H}_1$ and ${\cal H}_2$ denote the input and
output Hilbert spaces of $\cal E$.  
A density matrix can be regarded as an operator acting on the Hilbert 
space.   
\footnote{Even though the density matrix can be viewed as an operator,
it is important to remember that it represents a state and thus
evolves as a state rather than an operator.  For this reason we use
the name density matrix rather than the density operator. }
Let ${\cal B}({\cal H}_i)$ denote the set of all bounded operators
acting on ${\cal H}_i$ for $i=1,2$.  We can consider ${\cal E}(M)$ for
any $M \in {\cal B}({\cal H}_1)$ without restricting the domain to
density matrices.
A mapping $\cal E$ from ${\cal B}({\cal H}_1)$ to ${\cal B}({\cal
H}_2)$ is a quantum operation if it satisfies the following equivalent
sets of conditions:
\begin{enumerate} 
\item ${\cal E}$ is (i) linear, (ii) trace non-increasing ($\tr({\cal
E}(M)) \leq \tr(M)$) for all $M \geq 0$, and (iii) {\em completely
positive}.
The mapping ${\cal E}$ is called {\em positive} if ${M \geq
0}$ in ${\cal B}({\cal H}_1)$ implies ${\cal E}(M) \geq 0$ in 
${\cal B}({\cal H}_2)$. 
It is called completely positive if, for any auxillary Hilbert space
${\cal H}_a$, $\tilde M \geq 0$ in ${\cal B}({\cal H}_1 \otimes
{\cal H}_a$) implies $({\cal E} \ot {\cal I})(\tilde{M}) \geq 0$ 
in ${\cal B}({\cal H}_2 \otimes {\cal H}_a)$ where ${\cal I}_a$ is the
identity operation on ${\cal B}({\cal H}_a)$.
\item ${\cal E}$ has a {\em Kraus representation} or an {\em operator sum 
representation}~\cite{Schumacher96a,Choi75a,Kraus83a}:  
\bea
	{\cal E}(M) = \sum_k A_k M A^\dagger_k
\label{eq:osr_main}
\eea
where $\sum_k A^\dagger_k A_k \leq I$, and $I$ is the identity operator in 
${\cal B}({\cal H}_1)$.    
The $A_k$ operators are called the Kraus operators or the operation
elements of ${\cal E}$.
\item ${\cal E}(M) = \mbox{Tr}_o \lbm 
U (M \otimes \rho_a) U^\dagger (I \otimes P_o) \rbm$. 
Here, $\rho_a \in {\cal B}({\cal H}_a)$ is a density matrix of the
initial state of the ancilla, $I \in {\cal B}({\cal H}_2)$ is the
identity operator, ${\cal H}_2 \otimes {\cal H}_o = {\cal H}_1 \otimes
{\cal H}_a$, $P_o \in {\cal B}({\cal H}_o)$ is a projector, and
$\mbox{Tr}_o$ is a partial tracing over ${\cal H}_o$.
\end{enumerate}

Each set of conditions represents an approach to quantum operation 
when the input is a density matrix ($M = \rho$).
The first approach puts down three axioms any quantum operation should
satisfy.  The completely positive requirement states that if the
input is entangled with some other system (described by the Hilbert
space ${\cal H}_a$), the output after ${\cal E}$ acts on ${\cal H}_1$
should still be a valid state.
The third approach describes system-ancilla (or environment)
interaction.  Each of these evolutions results from a unitary
interaction of the system with a fixed ancilla state $\rho_a$,
followed by a measurement on a subsystem ${\cal H}_o$ with measurement
operators $\{P_o, I-P_o\}$, post-selection of the first outcome, and
removing ${\cal H}_o$.
\begin{figure}[http]

\setlength{\unitlength}{0.5mm}
\centering

\begin{picture}(170,50)

\put(40,35){\makebox(10,10){$\rho$ \Large \{ }}
\put(40,15){\makebox(10,10){$\rho_a$ \Large \{ }}
\put(120,40){\makebox(10,10){${\cal E}(\rho)$}}

\put(70,10){\framebox(25,40){\Huge{\em U}\hspace*{0.5ex}}}

\put(50,15){\line(1,0){20}}
\put(50,25){\line(1,0){20}}
\put(50,35){\line(1,0){20}}
\put(50,45){\line(1,0){20}}

\put(95,15){\line(1,0){10}}
\put(95,23){\line(1,0){20}}
\put(95,31){\line(1,0){20}}
\put(95,45){\line(1,0){20}}

\put(105,11){\framebox(14,8){$P_o?$}}

\put(5,37){\makebox{``System''}}
\put(5,17){\makebox{``Ancilla''}}
\put(135,42){\makebox{``Output''}}
\put(115,20){\makebox{$\left. \begin{array}{c}\\[8mm] \end{array} \right\}$ 
~~``Discard''}}

\end{picture}
\end{figure}
The fact that the third approach is equivalence to the first is
nontrivial -- the evolutions described by the third approach are
actually all the mappings that satisfy the three basic axioms.
Finally, the second approach provides a convenient representation 
useful in quantum information theory, particularly in quantum
error correction (see \cite{Nielsen00} for a review).

Proofs of the equivalence of the three approaches are summarized in
\cite{Nielsen98,Leung00t}.
There are four major steps, showing that the 1st set of conditions  
implies the 2nd set and vice versa, and similarly for the 2nd and 3rd
sets of conditions.
The most nontrivial step is to show that every linear and completely
positive map has a Kraus representation, and a proof due to
Choi~\cite{Choi75a} for the finite dimensional case will be described
next.

\section{Choi's proof}
\label{sec:proof}

The precise statement to be proved is that, if ${\cal E}$ is a
completely positive linear map from ${\cal B}({\cal H}_1)$ to ${\cal
B}({\cal H}_2)$, then ${\cal E}(M) = \sum_k A_k M A_k^\dagger$ for some
$n_2 \times n_1$ matrices $A_k$, where $n_i$ is the dimension of
${\cal H}_i$.
Let $|\Phi\> = \frac{1}{\sqrt{n_1}} \sum_i |i\> \otimes |i\>$ 
be a maximally entangled state in ${\cal H}_1 \otimes {\cal H}_1$.  
Here, $\{|i\>\}_{i=1,\cdots,n_1}$ is a basis for ${\cal H}_1$.
Consider $({\cal I} \otimes {\cal E})(\tilde{M})$ where 
\bea
	\tilde{M} =  n_1 |\Phi\>\<\Phi| = 
	\sum_{i,j=1}^{n_1} |i\>\<j| \otimes |i\>\<j| 
\,.
\eea
$\tilde{M}$ is an $n_1 \times n_1$ array of $n_1 \times n_1$ matrices.  
The $(i,j)$ block is exactly $|i\>\<j|$:   
\bea
M ~ = ~
{\footnotesize 
\left[ \begin{array}{c|c|c|c}  
\begin{array}{cccc} 
	1&0&\cdot&0 \\
	0&0&\cdot&0\\ 
	\cdot&\cdot&\cdot&\cdot \\
	0&0&\cdot&0 
\end{array}	
& 
\begin{array}{cccc} 
	0&1&\cdot&0 \\
	0&0&\cdot&0\\ 
	\cdot&\cdot&\cdot&\cdot \\
	0&0&\cdot&0 
\end{array}	
& 
\begin{array}{cccc} 
	\cdot&\cdot&\cdot&\cdot \\
	\cdot&\cdot&\cdot&\cdot \\
	\cdot&\cdot&\cdot&\cdot \\
	\cdot&\cdot&\cdot&\cdot 
\end{array}	
& 
\begin{array}{cccc} 
	0&0&\cdot&1 \\
	0&0&\cdot&0\\ 
	\cdot&\cdot&\cdot&\cdot \\
	0&0&\cdot&0 
\end{array}	
\\
\hline
\begin{array}{cccc} 	
	0&0&\cdot&0 \\
	1&0&\cdot&0\\ 
	\cdot&\cdot&\cdot&\cdot \\
	0&0&\cdot&0 
\end{array}	
& 
\begin{array}{cccc} 
	0&0&\cdot&0 \\
	0&1&\cdot&0\\ 
	\cdot&\cdot&\cdot&\cdot \\
	0&0&\cdot&0 
\end{array}	
& 
\begin{array}{cccc} 
	\cdot&\cdot&\cdot&\cdot \\
	\cdot&\cdot&\cdot&\cdot \\
	\cdot&\cdot&\cdot&\cdot \\
	\cdot&\cdot&\cdot&\cdot 
\end{array}	
& 
\begin{array}{cccc} 
	0&0&\cdot&0 \\
	0&0&\cdot&1 \\ 
	\cdot&\cdot&\cdot&\cdot \\
	0&0&\cdot&0 
\end{array}	
\\ 
\hline
	\begin{array}{cccc} 
	\cdot&\cdot&\cdot&\cdot \\
	\cdot&\cdot&\cdot&\cdot \\
	\cdot&\cdot&\cdot&\cdot \\
	\cdot&\cdot&\cdot&\cdot 
\end{array}	
& 
\begin{array}{cccc} 
	\cdot&\cdot&\cdot&\cdot \\
	\cdot&\cdot&\cdot&\cdot \\
	\cdot&\cdot&\cdot&\cdot \\
	\cdot&\cdot&\cdot&\cdot 
\end{array}	
& 
\begin{array}{cccc} 
	\cdot&\cdot&\cdot&\cdot \\
	\cdot&\cdot&\cdot&\cdot \\
	\cdot&\cdot&\cdot&\cdot \\
	\cdot&\cdot&\cdot&\cdot 
\end{array}	
& 
\begin{array}{cccc} 
	\cdot&\cdot&\cdot&\cdot \\
	\cdot&\cdot&\cdot&\cdot \\
	\cdot&\cdot&\cdot&\cdot \\
	\cdot&\cdot&\cdot&\cdot 
\end{array}	
\\
\hline
\begin{array}{cccc} 
	0&0&0&0 \\
	0&\cdot&\cdot&\cdot\\ 
	0&0&0&0 \\
	1&0&0&0 
\end{array}	
& 
\begin{array}{cccc} 
	0&0&0&0 \\
	0&\cdot&\cdot&\cdot\\ 
	0&0&0&0 \\
	0&1&0&0 
\end{array}	
& 
\begin{array}{cccc} 
	\cdot&\cdot&\cdot&\cdot \\
	\cdot&\cdot&\cdot&\cdot \\
	\cdot&\cdot&\cdot&\cdot \\
	\cdot&\cdot&\cdot&\cdot 
\end{array}	
& 
\begin{array}{cccc} 
	0&0&0&0 \\
	0&\cdot&\cdot&\cdot\\ 
	0&0&0&0 \\
	0&0&0&1 
\end{array}	
		   \end{array} \right]	
}
\label{eq:proof41}
\eea
When ${\cal I} \otimes {\cal E}$ is applied to $\tilde{M}$, 
the $(i,j)$ block becomes ${\cal E}(|i\>\<j|)$, and 
\bea
{\footnotesize 
({\cal I} \otimes {\cal E})(\tilde{M}) = \left[  \begin{array}{c|c|c|c}  
{\cal E} \left( 
\begin{array}{cccc} 
	1&0&\cdot&0 \\
	0&0&\cdot&0\\ 
	\cdot&\cdot&\cdot&\cdot \\
	0&0&\cdot&0 
\end{array}	
\right)
& 
{\cal E} \left( 
\begin{array}{cccc} 
	0&1&\cdot&0 \\
	0&0&\cdot&0\\ 
	\cdot&\cdot&\cdot&\cdot \\
	0&0&\cdot&0 
\end{array}	
\right)
& 
\begin{array}{cccc} 
	\cdot&\cdot&\cdot&\cdot \\
	\cdot&\cdot&\cdot&\cdot \\
	\cdot&\cdot&\cdot&\cdot \\
	\cdot&\cdot&\cdot&\cdot 
\end{array}	
& 
{\cal E} \left( 
\begin{array}{cccc} 
	0&0&\cdot&1 \\
	0&0&\cdot&0\\ 
	\cdot&\cdot&\cdot&\cdot \\
	0&0&\cdot&0 
\end{array}	
\right)
\\
\hline
{\cal E} \left( 
\begin{array}{cccc} 	
	0&0&\cdot&0 \\
	1&0&\cdot&0\\ 
	\cdot&\cdot&\cdot&\cdot \\
	0&0&\cdot&0 
\end{array}	
\right)
& 
{\cal E} \left( 
\begin{array}{cccc} 
	0&0&\cdot&0 \\
	0&1&\cdot&0\\ 
	\cdot&\cdot&\cdot&\cdot \\
	0&0&\cdot&0 
\end{array}	
\right)
& 
\begin{array}{cccc} 
	\cdot&\cdot&\cdot&\cdot \\
	\cdot&\cdot&\cdot&\cdot \\
	\cdot&\cdot&\cdot&\cdot \\
	\cdot&\cdot&\cdot&\cdot 
\end{array}	
& 
{\cal E} \left( 
\begin{array}{cccc} 
	0&0&\cdot&0 \\
	0&0&\cdot&1 \\ 
	\cdot&\cdot&\cdot&\cdot \\
	0&0&\cdot&0 
\end{array}	
\right)
\\ 
\hline
	\begin{array}{cccc} 
	\cdot&\cdot&\cdot&\cdot \\
	\cdot&\cdot&\cdot&\cdot \\
	\cdot&\cdot&\cdot&\cdot \\
	\cdot&\cdot&\cdot&\cdot 
\end{array}	
& 
\begin{array}{cccc} 
	\cdot&\cdot&\cdot&\cdot \\
	\cdot&\cdot&\cdot&\cdot \\
	\cdot&\cdot&\cdot&\cdot \\
	\cdot&\cdot&\cdot&\cdot 
\end{array}	
& 
\begin{array}{cccc} 
	\cdot&\cdot&\cdot&\cdot \\
	\cdot&\cdot&\cdot&\cdot \\
	\cdot&\cdot&\cdot&\cdot \\
	\cdot&\cdot&\cdot&\cdot 
\end{array}	
& 
\begin{array}{cccc} 
	\cdot&\cdot&\cdot&\cdot \\
	\cdot&\cdot&\cdot&\cdot \\
	\cdot&\cdot&\cdot&\cdot \\
	\cdot&\cdot&\cdot&\cdot 
\end{array}	
\\
\hline
{\cal E} \left( 
\begin{array}{cccc} 
	0&0&0&0 \\
	0&\cdot&\cdot&\cdot\\ 
	0&0&0&0 \\
	1&0&0&0 
\end{array}	
\right)
& 
{\cal E} \left( 
\begin{array}{cccc} 
	0&0&0&0 \\
	0&\cdot&\cdot&\cdot\\ 
	0&0&0&0 \\
	0&1&0&0 
\end{array}	
\right)
& 
\begin{array}{cccc} 
	\cdot&\cdot&\cdot&\cdot \\
	\cdot&\cdot&\cdot&\cdot \\
	\cdot&\cdot&\cdot&\cdot \\
	\cdot&\cdot&\cdot&\cdot 
\end{array}	
& 
{\cal E} \left( 
\begin{array}{cccc} 
	0&0&0&0 \\
	0&\cdot&\cdot&\cdot\\ 
	0&0&0&0 \\
	0&0&0&1 
\end{array}	
\right)
		   \end{array} \right]	
}
\label{eq:proof42}
\eea
which is an $n_1 \times n_1$ array of $n_2 \times n_2$ matrices.  

We now express $({\cal I} \otimes {\cal E})(\tilde{M})$ in a manner
completely independent of \eq{proof42}.  Since $\tilde{M}$ is positive
and ${\cal E}$ is completely positive, $({\cal I} \otimes {\cal
E})(\tilde{M})$ is positive, and can be expressed as $({\cal I}
\otimes {\cal E})(\tilde{M}) = \sum_{l = 1}^{n_1 n_2} |a_k\>\<a_k|$, 
where $|a_k\>$ are the eigenvectors of $({\cal I} \otimes {\cal
E})(\tilde{M})$, normalized to the respective eigenvalues.  One can
represent each $|a_k\>$ as a column vector and each $\<a_k|$ as a
row vector.  
We can divide the column vector $|a_k\>$ into $n_1$ segments each of
length $n_2$, and define a matrix $A_k$ with the $i$-th column being
the $i$-th segment, so that the $i$-th segment is exactly $A_k |i\>$.
Then 
\begin{figure}[http]
\setlength{\unitlength}{0.6mm}
\centering
\begin{picture}(210,130)

\put(2,125){\makebox{$({\cal I} \otimes {\cal E})(\tilde{M})=\sum_k $}}
\put(57,10){\framebox(8,120)}
\put(57,40){\line(1,0){8}}
\put(57,70){\line(1,0){8}}
\put(57,100){\line(1,0){8}}
\put(65,120){\makebox(10,10){$\times$}}
\put(75,122){\framebox(120,8)}
\put(105,122){\line(0,1){8}}
\put(135,122){\line(0,1){8}}
\put(165,122){\line(0,1){8}}
\put(72,25){\line(-1,0){10}}
\put(72,85){\line(-1,0){10}}
\put(72,105){\line(-1,0){10}}
\put(68,20){\makebox(25,10){$~ A_k|n_1\>$}}
\put(67,80){\makebox(25,10){$~ A_k|2\>$}}
\put(67,100){\makebox(25,10){$~ A_k|1\>$}}
\put(77,121){\makebox(25,10){$\<1|A_k^\dagger$}}
\put(108,121){\makebox(25,10){$\<2|A_k^\dagger$}}
\put(167,121){\makebox(25,10){$\<n_1|A_k^\dagger$}}

\end{picture}
\vspace*{-2ex}
\label{fig:proof43}
\end{figure}

and
\bea
({\cal I} \otimes {\cal E})(\tilde{M}) = \sum_k
{\footnotesize 
\left[ \begin{array}{c|c|c|c}  
\begin{array}{c} ~\\A_k |1\>\<1| A_k^\dagger\\~ \end{array}
& 
A_k |1\>\<2| A_k^\dagger
& 
~~~~\cdots~~~~
& 
A_k |1\>\<n_1| A_k^\dagger
\\
\hline
\begin{array}{c} ~\\A_k |2\>\<1| A_k^\dagger\\~ \end{array}
& 
A_k |2\>\<2| A_k^\dagger
& 
\cdots
& 
A_k |2\>\<n_1| A_k^\dagger
\\
\hline
\begin{array}{c} ~\\ \cdots \\~ \end{array}
& 
\cdots
& 
\cdots
& 
\cdots
\\
\hline
\begin{array}{c} ~\\A_k |n_1\>\<1| A_k^\dagger\\~ \end{array}
& 
A_k |n_1\>\<2| A_k^\dagger
& 
\cdots
& 
A_k |n_1\>\<n_1| A_k^\dagger
\end{array} \right]	
}
\label{eq:proof44}
\eea
Comparing Eqs.~(\ref{eq:proof42}) and (\ref{eq:proof44}) block by
block ${\cal E}(M) = \sum_k A_k M A_k^\dagger$ for $\forall M =
|i\>\<j|$, and thus $\forall M \in {\cal B}({\cal H}_1)$ by
linearity.

\section{Recipe for quantum process tomography}
\label{sec:qpt}

The basic assumptions in quantum process tomography are as follows.
The unknown quantum operation, ${\cal E}$, is given as an ``oracle''
or a ``blackbox'' one can call without knowing its internal
mechanism.
One prepares certain input states and {\em measures} the 
corresponding output density matrices to learn about ${\cal E}$ 
systematically.  
The task to measure the density matrix of a quantum system is called 
quantum state tomography~\cite{ADD}.  
To obtain a Kraus representation for ${\cal E}$, one needs an 
experimental procedure that specifies the input states to be prepared, 
and a numerical method for obtaining the Kraus operators from the 
measured output density matrices. 

A method follows immediately from the proof in Sec.~\ref{sec:proof}. 
We retain all the previously defined notations.
The crucial observation is that ${1 \over n_1} \tilde{M}$ and ${1
\over n_1}({\cal I} \otimes {\cal E})(\tilde{M})$ correspond to the
input and output physical states $|\Phi\>\<\Phi|$ and $({\cal I}
\otimes {\cal E})(|\Phi\>\<\Phi|)$ which can be prepared and measured.
The procedure is therefore to:
\begin{enumerate}
\item Prepare a maximally entangled 
state $|\Phi\>$ in ${\cal H}_1 \otimes {\cal H}_1$.
\item Subject one system to the action of ${\cal E}$, while making sure 
that the other system does not evolve.  
\item Measure the joint output density matrix $({\cal I} \otimes {\cal
E})(|\Phi\>\<\Phi|) = {1 \over n_1} ({\cal I} \otimes {\cal
E})(\tilde{M})$, multiply by $n_1$, obtain the eigen-decomposition $\sum_k
|a_k\>\<a_k|$.  Divide $|a_k\>$ (of length $n_1 n_2$) into $n_1$ equal
segments each of length $n_2$.  $A_k$ is the $n_2 \times n_1$ matrix
having the $i$-th segment as its $i$-th column.
\end{enumerate} 

The maximally entangled state in the above procedure can be replaced
by any pure state with maximum Schmidt number, $|\phi\> = \sum_i
\alpha_i (U |i\>) \otimes (V |i\>)$ where $\alpha_i \geq 0$ are real
and $\sum_i \alpha_i^2 = 1$.
The output density matrix $\rho_{\rm out}$ is equal to $({\cal I}
\otimes {\cal E})(|\phi\>\<\phi|) = \sum_{i,j} \alpha_i \alpha_j (U
|i\>\<j| U^\dagger) \otimes {\cal E}(V |i\>\<j| V^\dagger)$.  One
computes $(U^\dagger \otimes I) \, \rho_{\rm out} \, (U \otimes I)$,
divides the $(i,j)$ block by $\alpha_i \alpha_j$, and performs 
eigen-decomposition to obtain a set of $A_k$ operators.  The
Kraus operators of $\cal E$ are given by $A_k V^\dagger$.

\section{Discussion}
\label{sec:conclude}

We have provided an experimental and analytic procedure for obtaining
a set of Kraus operators $A_k$ for an unknown quantum operation.  The
set of $A_k$ is called ``canonical'' in \cite{Choi75a}, meaning that
the $A_k$ are linearly independent.
We remark that any other Kraus representation can be obtained from
$A_k$ using the fact that ${\cal E}(\rho) = \sum_k A_k \rho
A_k^\dagger = \sum_k B_k \rho B_k^\dagger$ if and only if $A_k =
\sum_j u_{kj} B_k$ when $u_{kj}$ are the entries of an isometry
\cite{Choi75a}.
Alternatively, one can replace the eigen-decomposition of 
$({\cal I} \otimes {\cal E})(|\Phi\>\<\Phi|)$ by any decomposition 
into a positive sum to obtain other valid sets of Kraus operators. 

Previous methods of quantum process tomography
\cite{Poyatos97a,Chuang96d,DAriano98} involve preparing a set of
physical input states $\rho_i$ that form a basis of ${\cal B}({\cal
H}_1)$, and measuring ${\cal E}(\rho_i)$ to determine ${\cal E}$.
The input states $\rho_i$ are physical states, and cannot be chosen to
be trace orthonormal, causing complications in the analysis.  
In contrast, the output state in the current method automatically
contains complete information on ${\cal E}(|i\>\<j|)$ for the
unphysical orthonormal basis $|i\>\<j|$ (see \eq{proof42}), which
greatly simplifies the analysis to obtain the Kraus operators.
However, the current method requires the preparation of a
maximally entangled state and the ability to stop the evolution  
of the reference system while $\cal E$ is being applied.
The previous methods are more suitable in implementations such as
solution NMR systems, while the current method is more suitable for
implementations such as optical systems.

Any efficient quantum process tomography procedure consumes
approximately the same amount of resources, which is determined by the
number of degrees of freedom in the quantum operation.  In general, to
measure an $n \times n$ density matrix, $n^2$ {\em ensemble}
measurements are needed, requiring $\approx {\cal O}(n^2)$ steps.  The
previous methods require the determination of $n_1^2$ density matrices
each $n_2 \times n_2$ and take $\approx {\cal O}(n_1^2 n_2^2)$ steps.
The current method requires the determination of one $n_1 n_2 \times
n_1 n_2$ density matrix which also requires $\approx {\cal O}(n_1^2
n_2^2)$ steps.  In both cases, the number of steps is of the same
order as the number of degrees of freedom in the quantum operation and
are optimal in some sense.

\section{Acknowledgement}

We thank Isaac Chuang for suggesting the application of Choi's proof
in quantum process tomography.  After the initial report of the
current result in \cite{Leung00t}, G. D'Ariano and P. Presti
independently reported a similar tomography method~\cite{DAriano00}.
This work is supported in part by the NSA and ARDA under the US Army
Research Office, grant DAAG55-98-C-0041


\end{document}